\documentclass[a4paper]{jpconf}
\usepackage{graphicx}
\usepackage{amsmath}
\usepackage{xcolor}
\definecolor{ballblue}{rgb}{0.13, 0.67, 0.8}
\definecolor{bluemunsell}{rgb}{0.0, 0.5, 0.69}
\usepackage{float}
\usepackage{adjustbox}
\usepackage[colorlinks,linkcolor = bluemunsell,
urlcolor  = bluemunsell,
citecolor = bluemunsell,
anchorcolor = bluemunsell]{hyperref}\usepackage{hyperref}
\usepackage{cite}
\newcommand*{\email}[1]{%
	\footnotesize\href{mailto:#1}{#1}}
\begin{document}
\title{Search for charged Higgs boson in $\mu\nu$ channel within 2HDM type-III}

\author{Rachid Benbrik$^1${\color{bluemunsell}{$^*$}}, Mohammed Boukidi$^1${\color{bluemunsell}$^\dagger$}, Bouzid Manaut$^2$, Mohammed Ouchemhou$^1$, Souad Semlali$^3$ and Souad Taj$^2$}

\address{$^1$Polydisciplinary Faculty, Laboratory of Fundamental and Applied Physics, Cadi Ayyad University, Sidi Bouzid, B.P. 4162, Safi, Morocco.\\
$^2$ Polydisciplinary Faculty, Laboratory of Research in Physics and Engineering Sciences,
Team of Modern and Applied Physics, Sultan Moulay Slimane University, Beni Mellal 23000, Morocco.\\$^3$ School of Physics and Astronomy, University of Southampton, Southampton, SO17 1BJ, United Kingdom.}

\ead{{\color{bluemunsell}$^*$}\email{r.benbrik@uca.ac.ma}, {\color{bluemunsell}$^\dagger$}\email{mohammed.boukidi@ced.uca.ma}}

\begin{abstract}
We discus the muonic decay of light charged Higgs boson for $m_{H^\pm}\le m_t-m_b$ in the context of the generic two Higgs doublet model (2HDM) type-III. We show that for both alignment limit scenarios, the $H^\pm\to\mu\nu$ signal could be dominant and give alternative
reachable signatures to those arising from top quark production and decay.
\end{abstract}

\section{Introduction}
After the discovery of a scalar particle at the Large Hadron Collider (LHC) \cite{Aad:2012tfa,Chatrchyan:2012ufa}, the Standard Model (SM) is confirmed to be a self consistent theory, However, despite its high compatibility  with the experimental measurements, the SM fails to answer some crucial questions such as  hierarchy problem, neutrino masses, gravity, etc. In this regard, any extension of the SM is well motivated. 

The Two-Higgs-Doublet Model (2HDM) is one of the simplest option, among the extended Higgs sector models.
The general $\mathcal{CP}$-conserving 2HDM contains five physical eigenstates: two $\mathcal{CP}$-even neutral scalars, where one of theme can be identified as SM-like with mass at 125 GeV, one $\mathcal{CP}$-odd Higgs
boson $A$, and a pair of charged Higgs boson $H^\pm$. 
The charged Higgs bosons presents a special challenge for experimental searches. They are dominantly produced in association with top quarks ($tbH^\pm$), the decay mode $H^\pm \to \mu \nu $ can hold a chance for discovering light charged Higgses at the LHC. 

The purpose of this contribution is to explore $pp \to t b H^\pm$, $H^\pm \to \mu \nu $ in the current context of LHC, to assess the extent to which they might complement the searches for light charged Higgs.  We will demonstrate that the production rates of such alternative production channel, in type-III, have the potential to be overwhelmingly stronger than the production channels followed $H^\pm \to \tau \nu/tb$.  More specifically, we will illustrate that the muonic decay of a lightly charged Higgs boson, could be dominant and give alternative reachable signatures to those arising from top quark production and decay. Consequently, such modes can serve as new discovery channels for light $H^\pm$ states at the LHC.

The contribution is organized as follows: In the first Section we shall introduce some basic notation of 2HDM, and  in section 2 we review the most important
constraints. Then, in the following sections we will discuss the numerical results. We will finally
conclude. 
\section{Review of 2HDM Type-III}
%
In the Yukawa sector, the most general scalar to fermions couplings are expressed by:
\begin{eqnarray}
	-{\cal L}_Y &=& \bar Q_L Y^u_1 U_R \tilde \Phi_1 + \bar Q_L Y^{u}_2 U_R
	\tilde \Phi_2  + \bar Q_L Y^d_1 D_R \Phi_1 
	+ \bar Q_L Y^{d}_2 D_R \Phi_2 \nonumber \\
	&+&  \bar L Y^\ell_1 \ell_R \Phi_1 + \bar L Y^{\ell}_2 \ell_R \Phi_2 + H.c. 
	\label{eq:Yu}
\end{eqnarray}
where $Q_L = (u_L , d_L )$ and  $L = (\ell_L , \nu_L )$  are the doublets of $S U(2)_L$ , and $Y^{f,\ell}_{1,2}$  denote the $3\times 3$ Yukawa matrices. 

In order to  keep the FCNCs under control, while inducing flavor violating Higgs signals, we adopt the description presented in \cite{Chen:2018ytc,Hernandez-Sanchez:2012vxa,Cheng:1987rs, Diaz-Cruz:2004wsi} by assuming a flavor symmetry that suggest a specific texture of the Yukawa matrices, where the non-diagonal Yukawa couplings, $\tilde{Y}_{ij}$, are given in terms of fermions masses and dimensionless real parameter, $\tilde{Y}_{ij} \propto \sqrt{m_i m_j}/ v ~\chi_{ij}$.\\
Therefore, after spontaneous symmetry breaking the Yukawa Lagrangian can be written, in terms of the mass eigenstates of the Higgs bosons, as follows:
\begin{align}
	-{\cal L}^{III}_Y  &= \sum_{f=u,d,\ell} \frac{m^f_j }{v} \times\left( (\xi^f_h)_{ij}  \bar f_{Li}  f_{Rj}  h + (\xi^f_H)_{ij} \bar f_{Li}  f_{Rj} H - i (\xi^f_A)_{ij} \bar f_{Li}  f_{Rj} A \right)\nonumber\\  &+ \frac{\sqrt{2}}{v} \sum_{k=1}^3 \bar u_{i} \left[ \left( m^u_i  (\xi^{u*}_A)_{ki}  V_{kj} P_L+ V_{ik}  (\xi^d_A)_{kj}  m^d_j P_R \right) \right] d_{j}  H^+ \nonumber\\  &+ \frac{\sqrt{2}}{v}  \bar \nu_i  (\xi^\ell_A)_{ij} m^\ell_j P_R \ell_j H^+ + H.c.\, \label{eq:Yukawa_CH}
\end{align} 
 The reduced Yukawa couplings $(\xi^{f,\ell}_\phi)_{ij}$ are given in Table~\ref{coupIII}, in terms of the free parameters\footnote{The free parameters $\chi_{ij}^{f,\ell}$ are tested at the current B  physics constraints (more details can be found in Refs \cite{Benbrik:2021wyl, Benbrik:2022azi}).} $\chi_{ij}^{f,\ell}$ and the mixing angle $\alpha$ and  of $\tan\beta$.
 \begin{table}[h!]
 	\begin{center}
 		\setlength{\tabcolsep}{11pt}
 		\renewcommand{\arraystretch}{0.6} %
 		\begin{tabular}{c|c|c|c} \hline\hline 
 			$\phi$  & $(\xi^u_{\phi})_{ij}$ &  $(\xi^d_{\phi})_{ij}$ &  $(\xi^\ell_{\phi})_{ij}$  \\   \hline
 			$h$~ 
 			& ~ $  \frac{c_\alpha}{s_\beta} \delta_{ij} -  \frac{c_{\beta-\alpha}}{\sqrt{2}s_\beta}  \sqrt{\frac{m^u_i}{m^u_j}} \chi^u_{ij}$~
 			& ~ $ -\frac{s_\alpha}{c_\beta} \delta_{ij} +  \frac{c_{\beta-\alpha}}{\sqrt{2}c_\beta} \sqrt{\frac{m^d_i}{m^d_j}}\chi^d_{ij}$~
 			& ~ $ -\frac{s_\alpha}{c_\beta} \delta_{ij} + \frac{c_{\beta-\alpha}}{\sqrt{2}c_\beta} \sqrt{\frac{m^\ell_i}{m^\ell_j}}  \chi^\ell_{ij}$ ~ \\
 			$H$~
 			& $ \frac{s_\alpha}{s_\beta} \delta_{ij} + \frac{s_{\beta-\alpha}}{\sqrt{2}s_\beta} \sqrt{\frac{m^u_i}{m^u_j}} \chi^u_{ij} $
 			& $ \frac{c_\alpha}{c_\beta} \delta_{ij} - \frac{s_{\beta-\alpha}}{\sqrt{2}c_\beta} \sqrt{\frac{m^d_i}{m^d_j}}\chi^d_{ij} $ 
 			& $ \frac{c_\alpha}{c_\beta} \delta_{ij} -  \frac{s_{\beta-\alpha}}{\sqrt{2}c_\beta} \sqrt{\frac{m^\ell_i}{m^\ell_j}}  \chi^\ell_{ij}$ \\
 			$A$~  
 			& $ \frac{1}{t_\beta} \delta_{ij}- \frac{1}{\sqrt{2}s_\beta} \sqrt{\frac{m^u_i}{m^u_j}} \chi^u_{ij} $  
 			& $ t_\beta \delta_{ij} - \frac{1}{\sqrt{2}c_\beta} \sqrt{\frac{m^d_i}{m^d_j}}\chi^d_{ij}$  
 			& $t_\beta \delta_{ij} -  \frac{1}{\sqrt{2}c_\beta} \sqrt{\frac{m^\ell_i}{m^\ell_j}}  \chi^\ell_{ij}$ \\ \hline \hline 
 		\end{tabular}
 	\end{center}
 	\caption {Yukawa couplings of the $h$, $H$, and $A$ bosons to the quarks and leptons in the 2HDM Type-III.} 
 	\label{coupIII}
 \end{table}

\section{Theoretical and experimental constraints}
In this section we list the constraints applied in our studies:
\begin{itemize}
	\item \textbf{Theoretical constraints}: included are Unitarity ~\cite{uni1,uni2,uni3}, Perturbativity~\cite{Branco:2011iw}, and Vacuum stability \cite{Barroso:2013awa,sta}. All these constraints are tested via the publicly
	available two-higgs doublet model calculator \texttt{2HDMC-1.8.0} tool \cite{2hdmc}. 
   \item  \textbf{Collider constraints}: included are agreement with electroweak precision observables(EWPOs) \cite{Baak:2014ora} through the oblique parameters ($S$, $T$, $U$) ~\cite{Grimus:2007if,oblique2,Haller:2018nnx}, agreement  with current collider measurements of the Higgs signal strength as well as limits obtained from various searches of additional Higgs bosons at the LEP, Tevatron and LHC, in which we make use of  \texttt{HiggsBouns-5.9.0}\cite{HB} and  \texttt{HiggsSignal-2.6.0}\cite{HS}. And finally we ask for agreement with the current limits from B physics observables by using the public code \texttt{SuperIso-v4.1} \cite{superIso}.
 
\end{itemize}

\section{Results and discussion} 
\begin{figure}[H]	
	\includegraphics[height=7.5cm,width=15cm]{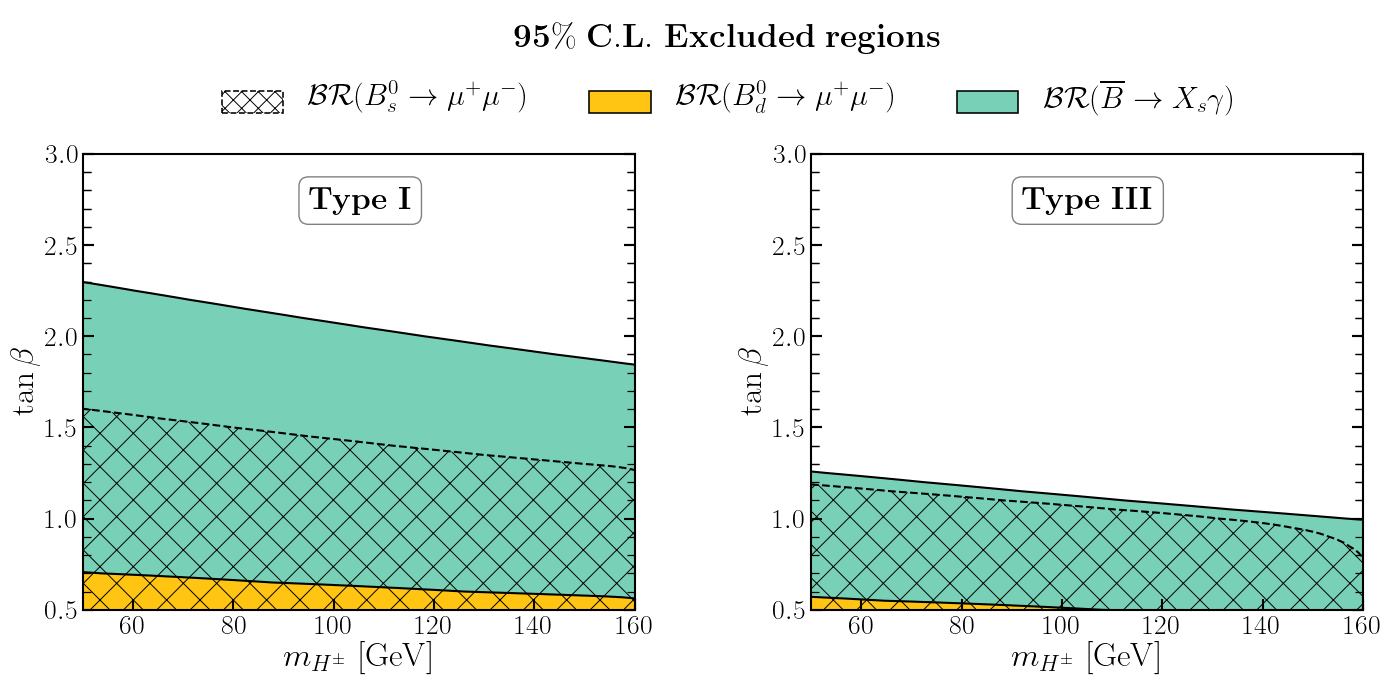}\hspace{8cm}
	\caption{Excluded regions of the $(m_{H^\pm},~\tan\beta)$ parameter space  by flavour constraints at $95\%$ C.L..}
	\label{fig1}	
\end{figure}

As a first check, on the B physics constraints, we show in Figure \ref{fig1}, the relevant constraints related to flavour observables: $B_{u}\to \tau\nu$, $B_{s,d}^{0}\to \mu^{+}\mu^{-}$ and $\bar{B}\to X_{s}\gamma$ which compatible the measurements at 95$\%$ C.L in  type-I (left) and type-III (right) panels.
 It is clear from Fig. \ref{fig1}  that the small values of $\tan\beta$ ($<1.3$) are excluded by flavour physics constraints for different masses of $H^\pm$ in type-III, unlike type-I. Hence the possibility for light charged Higgs together with small $\tan\beta$ still compatible with recent measurements in type-III.
 
\begin{figure}[H]	
		\centering
		\includegraphics[height=6cm,width=16cm]{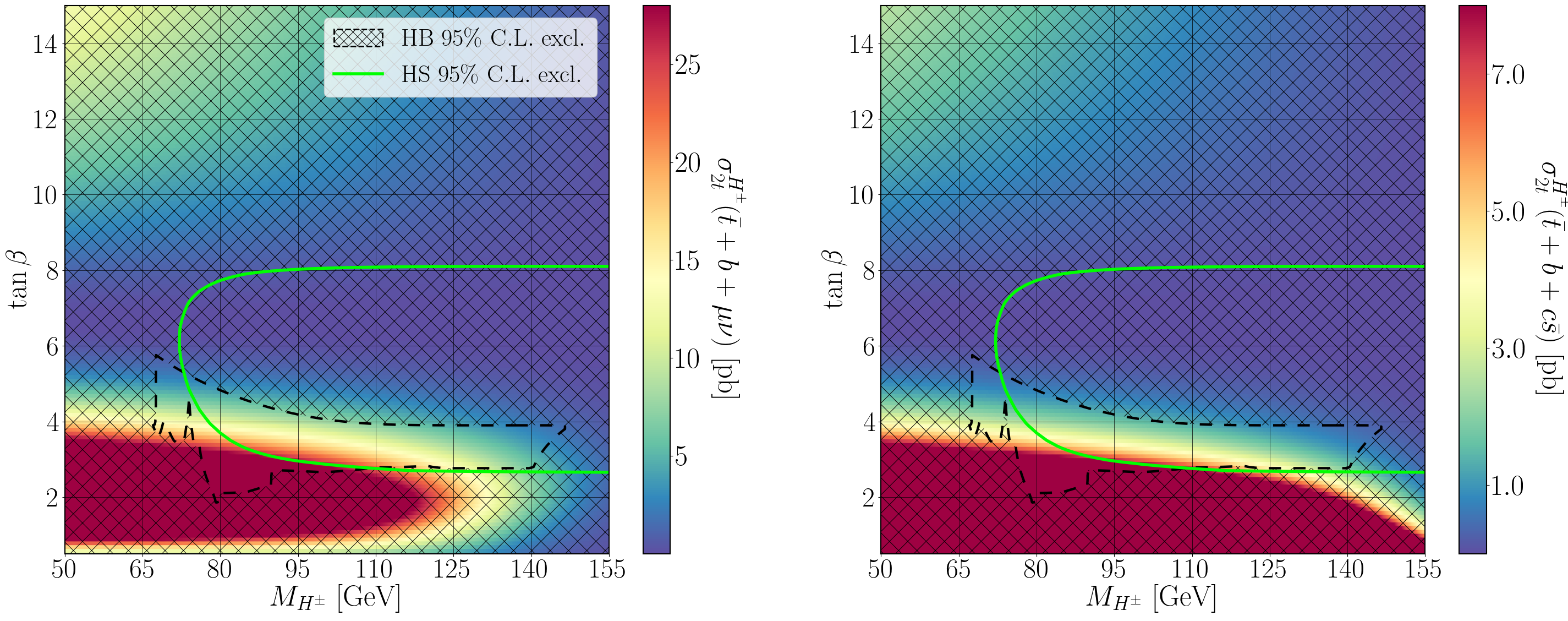}
	\caption{The  ${\sigma^{H^{\pm}}_{2t}(\bar{t}+b+XY)}$ mapped over the $(m_{H^{\pm}} , \tan\beta)$ plane. For $XY \equiv c\bar{s}$ (left) and  $XY \equiv \mu\nu $ (right). The hatched area  is excluded by the searches for
		additional Higgs bosons, while the solid green solid shows the exclusion limits
		from \texttt{HiggsSignals} at  2$\sigma$ C.L.  }
	\label{fig2}	
\end{figure}

\begin{figure}[H]	
	\centering
	\includegraphics[height=6cm,width=16cm]{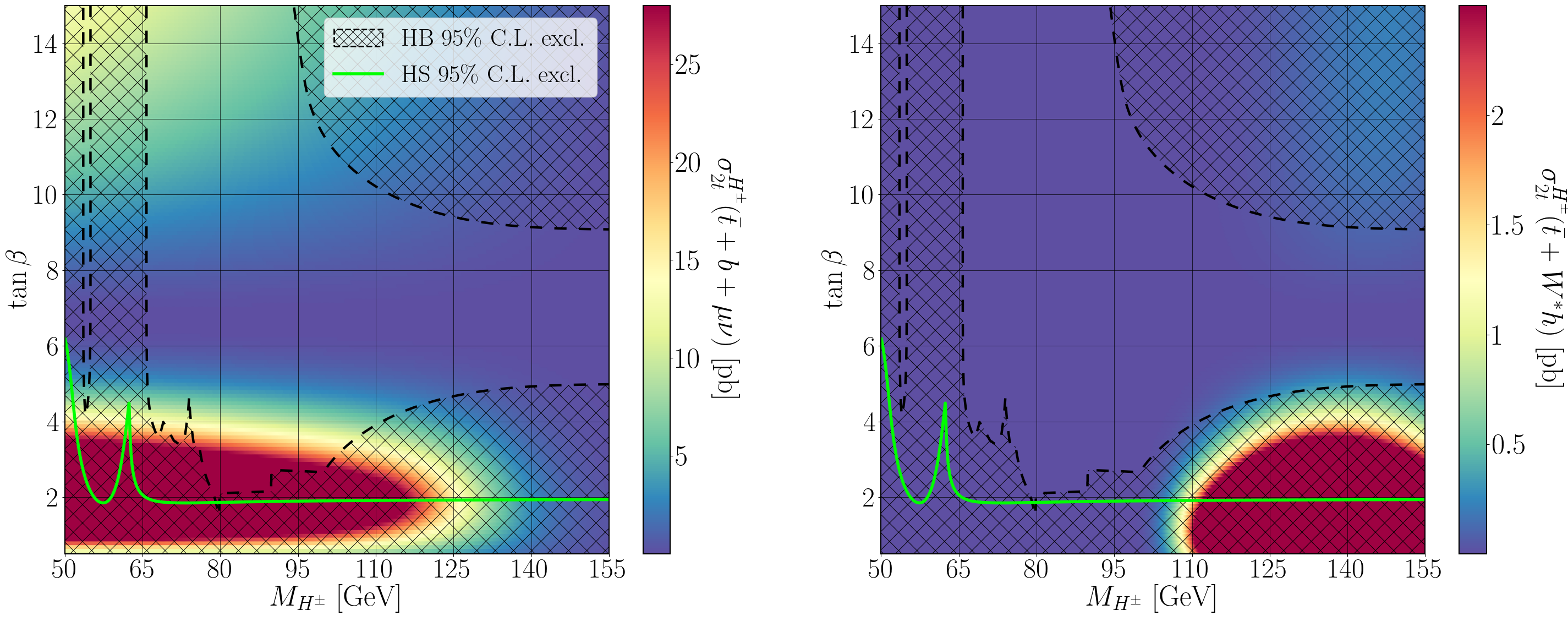}
	\caption{The  ${\sigma^{H^{\pm}}_{2t}(\bar{t}+b+XY)}$ mapped over the $(m_{H^{\pm}} , \tan\beta)$ plane. For $XY \equiv W^*h$ (right) and  $XY \equiv \mu\nu $ (left). The hatched area  is excluded by the searches for
		additional Higgs bosons, while the solid green solid shows the exclusion limits
		from \texttt{HiggsSignals} at  2$\sigma$ C.L. }
	\label{fig3}	
\end{figure}

  In what follows, we present the production rates of relevant final states of light charged Higgs for both scenarios inverted and standard hierarchy. In Fig.~\ref{fig2}, we show the production rates ${\sigma^{H^{\pm}}_{2t}(\bar{t}+b+\mu\nu)}$ (left) and  ${\sigma^{H^{\pm}}_{2t}(\bar{t}+b+c\bar{s})}$ (right) in the $(m_{H^\pm},\tan\beta)$ plane in the inverted hierarchy scenario. As in Fig.~\ref{fig3} (similar to Fig.~\ref{fig2}) we present ${\sigma^{H^{\pm}}_{2t}(\bar{t}+b+\mu\nu)}$ (left) and  ${\sigma^{H^{\pm}}_{2t}(\bar{t}+b+W^*h)}$ (right) in the $(m_{H^\pm},\tan\beta)$ in the standard hierarchy scenario. One can read from these figures that the signal cross section
  ${\sigma^{H^{\pm}}_{2t}(\bar{t}+b+\mu\nu)}$  can reach more than 23 pb in both scenarios. Therefore the  $H^\pm\to\mu\nu$  can be an excellent
 alternative discovery mode for light charged Higgs bosons at the LHC in the context of the 2HDM Type III.
\section*{Inverted hierarchy scenario}
In this scenario, we assume that the Higgs-like particle is $h$ with $m_h=125$ GeV. Then we perform a systematic scan over the 2HDM parameters using the following ranges:
\begin{eqnarray}
	\centering
	\centering
	&&\hspace{0.7cm} m_{h} = 125\ \text{GeV}, \ \ m_{H}= 135\ \text{GeV},\sin(\beta-\alpha)= -0.98, \ \  m_{A}= 220\ \text{GeV}, \nonumber \\
	&&\hspace{0.7cm} m_{H^\pm}\in [50,~160]\ \text{GeV}, \ \  \tan\beta\in [0.5,~15], \hspace{0.25cm}m_{12}^2 = m_h^2\tan\beta/(1+\tan^2\beta).
	\label{parm}
\end{eqnarray}

\section*{Standard hierarchy scenario}
In this scenario, we assume that the Higgs-like particle is $H$ with $m_H=125$ GeV, while $m_h$ is fixed at 95 GeV. Then we perform a systematic scan over the 2HDM parameters using the following ranges:
\begin{eqnarray}
	\centering
	&&\hspace{0.7cm} m_{h} = 95\ \text{GeV}, \ \ m_{H}= 125\ \text{GeV}, \sin(\beta-\alpha)= -0.05, \ \  m_{A}\in 177\ \text{GeV}, \nonumber \\
	&&\hspace{0.7cm} m_{H^\pm}\in [50,~160]\ \text{GeV}, \ \  \tan\beta\in [0.5,~15], \hspace{0.25cm}m_{12}^2 = m_h^2\tan\beta/(1+\tan^2\beta).
	\label{parm}
\end{eqnarray}

\section{Benchmark points}
In Table \ref{tab8} we present some Benchmark Points (BPs) for each scenario. All these BPs satisfy the most update theoretical and experimental constraints. For every BP in the table, we give the cross sections of $t+b+XY$, where $XY=cs, t^* b, W^{(*)}h, \tau\nu$ and $\mu\nu$ signatures. 
\begin{table}[H]
	\centering
			\begin{adjustbox}{max width=0.95\textwidth}
			\setlength{\tabcolsep}{30pt}
	\begin{tabular}{|c|c|c|c|c|c|}
		\hline
		\multicolumn{6}{|c|}{Selected BPs in scenario 1 (Inverted hierarchy)} \\\hline
		Parameters & BP1 & BP2   & BP3 & BP4 & BP5\\\hline
		\multicolumn{6}{|c|}{The Higgs masses are in GeV} \\\hline
		$m_h$ &  $125$ & $125$ & $125$ & $125$& $125$\\
		$m_H$ &  $135$ & $135$  & $135$ & $135$& $135$\\
		$m_A$ &  $220$ & $220$  & $220$& $220$ & $220$\\  
		$m_{H^\pm}$ &  $110$ & $93.2$  & $99.4$& $91.4$& $102$\\ 
		$\cos{(\beta-\alpha)}$ & $0.145$ & $0.145$ & $0.145$& $0.145$& $0.145$\\ 
		$\tan\beta$ & $2.9$ & $3.2$ & $3.7$& $4.3$& $3$ \\
		$m_{12}^2$ & $4815.35$ & $4448.39$ & $3935.5$& $3447.28$ & $4687.5$\\ \hline
		\multicolumn{6}{|c|}{${\sigma^{H^{\pm}}_{2t}(\bar{t}+b+XY)}$ [pb]} \\\hline
		$c\bar{s}$&$6.2$ &$5.76$ & $2.48$ & $1.32$& $6.48$\\
		$t^*\bar{b}$&$0.2$& $-$ &$0.002$& $-$& $0.04$\\
		$W^{(*)}h$&$-$&$-$&$-$&$-$&$-$\\
		$\mu\nu$&$21.12$&$11.46$&$22.92$&$6.82$&$23.32$\\
		$\tau\nu$&$2$&$1.36$&$0.34$&$0.08$&$1.88$\\
		\hline
		\multicolumn{6}{|c|}{Selected BPs in scenario 2 (Standard hierarchy)} \\\hline
		Parameters & BP1 & BP2   & BP3 & BP4 & BP5\\\hline
		\multicolumn{6}{|c|}{The Higgs masses are in GeV} \\\hline
		$m_h$ &  $95$ & $95$ & $95$& $95$ & $95$ \\
		$m_H$ &  $125$ & $125$  & $125$& $125$  & $125$ \\
		$m_A$ &  $177$ & $177$  & $177$& $177$  & $177$ \\  
		$m_{H^\pm}$ &  $95$ & $90$  & $94$  & $99.6$  & $90.80$\\ 
		$\sin{(\beta-\alpha)}$ & $-0.05$ & $-0.05$ & $-0.05$&$-0.05$ & $-0.05$\\ 
		$\tan\beta$ & $3$ & $3.5$ & $4.2$   & $3.7$ & $3.2$ \\
		$m_{12}^2$ & $2707.5$ & $2383.96$ & $2033.53$     & $2273.14$ & $2569.39$\\ \hline
		\multicolumn{6}{|c|}{${\sigma^{H^{\pm}}_{2t}(\bar{t}+b+XY)}$ [pb]} \\\hline
		$c\bar{s}$&$7.58$&$3.98$&$1.42$&$2.46$&$6.06$\\
		$t^*\bar{b}$&$-$&$-$&$-$&$0.003$&$-$\\
		$W^{(*)}h$&$-$&$-$&$-$&$0.00046$&$-$\\
		$\mu\nu$&$26.92$&$17.8$& $7.36$&$11.86$&$23.96$\\
		$\tau\nu$&$2.18$&$0.66$&$0.1$&$0.34$&$1.42$\\\hline	
	\end{tabular}	
\end{adjustbox}
	\caption{Mass spectra, mixing angles, and cross sections (in pb) in each configuration} \label{tab8}
\end{table}

\section{Conclusion}
In this contribution, we intended to explore the phenomenology of charged Higgs in the context of the generic 2HDM type-III. We have focused on the production of charged Higgs bosons via $pp\to t b H^\pm$ at the LHC with $\sqrt{s} = 13$ TeV. After considering all the updated theoretical and experimental constraints, we have studied the final states $t+b+\mu\nu$ as potential discovery channel over the traditional one $\tau\nu$. Finally to encourage experimentalists to look for the charged Higgs within this decay mode as a final state, we  proposed some BPs for both scenarios that are suitable for further experimental investigation. 

\section*{Acknowledgments}
This work is supported by the Moroccan Ministry of Higher Education and Scientific Research MESRSFC and CNRST Project PPR/2015/6.


\newpage
\section*{References}
\begin{thebibliography}{9}
\bibitem{Aad:2012tfa}
\textbf{ATLAS} collaboration, G.~Aad \underline{et al.}, ``Observation of a new particle in the search for the Standard Model Higgs boson with the ATLAS detector at the LHC'', \href{https://www.sciencedirect.com/science/article/pii/S037026931200857X}{\underline{Phys. Lett.} \textbf{B716} (2012) 1-29}, \href{https://arxiv.org/pdf/1207.7214.pdf}{\texttt{arXiv:1207.7214 [hep-ex]}}.\vspace{0.15cm}	

\bibitem{Chatrchyan:2012ufa}
\textbf{CMS} collaboration, S.~Chatrchyan \underline{et al.}, ``Observation of a New Boson at a Mass of 125 GeV with the CMS Experiment at the LHC'', \href{https://www.sciencedirect.com/science/article/pii/S0370269312008581}{\underline{Phys. Lett.} \textbf{B716} (2012) 30-61}, \href{https://arxiv.org/pdf/1207.7235.pdf}{\texttt{arXiv:1207.7235 [hep-ex]}}.\vspace{0.15cm}
\bibitem{Chen:2018ytc}
C.~H.~Chen and T.~Nomura,``$Re(\epsilon'_K/\epsilon_K$) and $K \to \pi \nu \bar\nu$ in a two-Higgs doublet model'', \href{https://link.springer.com/article/10.1007/JHEP08(2018)145}{\underline{JHEP} \textbf{08} (2018), 145}, \href{https://arxiv.org/pdf/1804.06017.pdf}{\texttt{arXiv:1804.06017}}.\vspace{0.15cm}

\bibitem{Hernandez-Sanchez:2012vxa}
J.~Hernandez-Sanchez, S.~Moretti, R.~Noriega-Papaqui and A.~Rosado,``Off-diagonal terms in Yukawa textures of the Type-III 2-Higgs doublet model and light charged Higgs boson phenomenology'', \href{https://link.springer.com/article/10.1007/JHEP07(2013)044}{\underline{JHEP}  \textbf{07} (2013), 044}, \href{https://arxiv.org/pdf/1212.6818.pdf}{\texttt{arXiv:1212.6818 [hep-ph]}}.\vspace{0.15cm}	

\bibitem{Cheng:1987rs}
T.~P.~Cheng and M.~Sher, ``Mass Matrix Ansatz and Flavor Nonconservation in Models with Multiple Higgs Doublets'', \href{https://journals.aps.org/prd/abstract/10.1103/PhysRevD.35.3484}{\underline{Phys. Rev.} \textbf{D35} (1987) 3484}.\vspace{0.15cm}

\bibitem{Diaz-Cruz:2004wsi}
J.~L.~Diaz-Cruz, R.~Noriega-Papaqui and A.~Rosado, ``Mass matrix ansatz and lepton flavor violation in the THDM-III'', \href{https://journals.aps.org/prd/abstract/10.1103/PhysRevD.69.095002}{\underline{Phys. Rev.} \textbf{D69} (2004) 095002}, \href{https://arxiv.org/pdf/hep-ph/0401194.pdf}{\texttt{arXiv:hep-ph/0401194 [hep-ph]}}.\vspace{0.15cm}

\bibitem{uni1}
S.~Kanemura, T.~Kubota and E.~Takasugi, ``Lee-Quigg-Thacker bounds for Higgs boson masses in a two doublet model'', \href{https://www.sciencedirect.com/science/article/abs/pii/0370269393912052?via/3Dihub}{\underline{Phys. Lett.} \textbf{B313} (1993) 155-160}, \href{https://arxiv.org/pdf/hep-ph/9303263.pdf}{\texttt{arXiv:hep-ph/9303263 [hep-ph]}}.\vspace{0.15cm}

\bibitem{uni2}
A.~G.~Akeroyd, A.~Arhrib and E.~M.~Naimi, ``Note on tree level unitarity in the general two Higgs doublet model'',  \href{https://www.sciencedirect.com/science/article/abs/pii/S037026930000962X?via/3Dihub}{\underline{Phys. Lett.} \textbf{B490} (2000) 119-124}, \href{https://arxiv.org/pdf/hep-ph/0006035.pdf}{\texttt{arXiv:hep-ph/0006035 [hep-ph]}}.\vspace{0.15cm}


\bibitem{uni3}
A.~Arhrib, ``Unitarity constraints on scalar parameters of the standard and two Higgs doublets model'', \href{https://arxiv.org/pdf/hep-ph/0012353.pdf}{\texttt{arXiv:hep-ph/0012353 [hep-ph]}}.\vspace{0.15cm}

\bibitem{Branco:2011iw}
G.~C.~Branco, P.~M.~Ferreira, L.~Lavoura, M.~N.~Rebelo, M.~Sher and J.~P.~Silva, ``Theory and phenomenology of two-Higgs-doublet models'', \href{https://www.sciencedirect.com/science/article/abs/pii/S0370157312000695?via/3Dihub}{\underline{Phys. Rept.} \textbf{516} (2012) 1-102}, \href{https://arxiv.org/pdf/1106.0034.pdf}{\texttt{arXiv:1106.0034 [hep-ph]}}.\vspace{0.15cm}

\bibitem{Barroso:2013awa}
A.~Barroso, P.~M.~Ferreira, I.~P.~Ivanov and R.~Santos, ``Metastability bounds on the two Higgs doublet model'', \href{https://doi.org/10.1007/JHEP06(2013)045}{\underline{JHEP} \textbf{06} (2013), 045}, \href{https://arxiv.org/pdf/1303.5098.pdf}{\texttt{arXiv:1303.5098 [hep-ph]}}.\vspace{0.15cm}


\bibitem{sta}
N.~G.~Deshpande and E.~Ma, ``Pattern of Symmetry Breaking with Two Higgs Doublets'', \href{https://journals.aps.org/prd/abstract/10.1103/PhysRevD.18.2574}{\underline{Phys. Rev.} \textbf{D18} (1978) 2574}.\vspace{0.15cm}

\bibitem{2hdmc}
D.~Eriksson, J.~Rathsman and O.~Stal, ``2HDMC: Two-Higgs-Doublet Model Calculator Physics and Manual,''
\href{https://www.sciencedirect.com/science/article/abs/pii/S0010465509003014?via/3Dihub}{\underline{Comput. Phys. Commun.} \textbf{181} (2010) 189-205}, \href{https://arxiv.org/pdf/0902.0851.pdf}{\texttt{arXiv:0902.0851 [hep-ph]}}.\vspace{0.15cm}

\bibitem{Benbrik:2021wyl}
R.~Benbrik, M.~Boukidi, B.~Manaut, M.~Ouchemhou, S.~Semlali and S.~Taj, ``New charged Higgs boson discovery channel at the LHC'', \href{https://arxiv.org/pdf/2112.07502.pdf}{\texttt{arXiv:2112.07502 [hep-ph]}}.\vspace{0.15cm}

\bibitem{Benbrik:2022azi}
R.~Benbrik, M.~Boukidi, S.~Moretti and S.~Semlali,
``Explaining the 96 GeV Di-photon Anomaly in a Generic 2HDM Type-III'', \href{https://arxiv.org/pdf/2204.07470.pdf}{\texttt{arXiv:2204.07470 [hep-ph]}}.\vspace{0.15cm}

\bibitem{Baak:2014ora}
M.~Baak \textit{et al.} [Gfitter Group], ``The global electroweak fit at NNLO and prospects for the LHC and ILC'', \href{https://link.springer.com/article/10.1140/epjc/s10052-014-3046-5}{\underline{Eur. Phys. J.}  \textbf{C74} (2014), 3046}, \href{https://arxiv.org/pdf/1407.3792.pdf}{\texttt{arXiv:1407.3792 [hep-ph]}}.\vspace{0.15cm}


\bibitem{Grimus:2007if}
W.~Grimus, L.~Lavoura, O.~M.~Ogreid and P.~Osland, ``A Precision constraint on multi-Higgs-doublet models'', \href{https://iopscience.iop.org/article/10.1088/0954-3899/35/7/075001}{\underline{J. Phys.}  \textbf{G35} (2008), 075001}, \href{https://arxiv.org/pdf/0711.4022.pdf}{\texttt{arXiv:0711.4022 [hep-ph]}}.\vspace{0.15cm}



\bibitem{oblique2}
W.~Grimus, L.~Lavoura, O.~M.~Ogreid and P.~Osland, ``The Oblique parameters in multi-Higgs-doublet models'', \href{https://www.sciencedirect.com/science/article/abs/pii/S0550321308002289?via/3Dihub}{\underline{Nucl. Phys.}  \textbf{B801} (2008) 81-96}, \href{https://arxiv.org/pdf/0802.4353.pdf}{\texttt{arXiv:0802.4353 [hep-ph]}}.\vspace{0.15cm}


\bibitem{Haller:2018nnx}
J.~Haller, A.~Hoecker, R.~Kogler, K.~M\"onig, T.~Peiffer and J.~Stelzer, ``Update of the global electroweak fit and constraints on two-Higgs-doublet models'', \href{https://doi.org/10.1140/epjc/s10052-018-6131-3}{\underline{Eur. Phys. J.} \textbf{C78} (2018) no.8, 675}, \href{https://arxiv.org/pdf/1803.01853.pdf}{\texttt{arXiv:1803.01853 [hep-ph]}}.\vspace{0.15cm}

\bibitem{HB}
P.~Bechtle, D.~Dercks, S.~Heinemeyer, T.~Klingl, T.~Stefaniak, G.~Weiglein and J.~Wittbrodt, ``HiggsBounds-5: Testing Higgs Sectors in the LHC 13 TeV Era'', \href{https://link.springer.com/article/10.1140/2Fepjc/2Fs10052-020-08557-9}{\underline{Eur. Phys. J.}  \textbf{C80} (2020) 1211}, \href{https://arxiv.org/pdf/2006.06007.pdf}{\texttt{arXiv:2006.06007 [hep-ph]}}.\vspace{0.15cm}

\bibitem{HS}
P.~Bechtle, S.~Heinemeyer, T.~Klingl, T.~Stefaniak, G.~Weiglein and J.~Wittbrodt, ``HiggsSignals-2: Probing new physics with precision Higgs measurements in the LHC 13 TeV era'', \href{https://link.springer.com/article/10.1140/2Fepjc/2Fs10052-021-08942-y}{\underline{Eur. Phys. J.}  \textbf{C81} (2021) 145},  \href{https://arxiv.org/pdf/2012.09197.pdf}{\texttt{arXiv:2012.09197 [hep-ph]}}.\vspace{0.15cm}

\bibitem{superIso}
F.~Mahmoudi, ``SuperIso v2.3: A Program for calculating flavor physics observables in Supersymmetry'', \href{https://www.sciencedirect.com/science/article/abs/pii/S0010465509000721?via/3Dihub}{\underline{Comput. Phys. Commun.} \textbf{180} (2009) 1579-1613}, \href{https://arxiv.org/pdf/0808.3144.pdf}{\texttt{arXiv:0808.3144 [hep-ph]}}.\vspace{0.15cm}
\end{thebibliography}
\end{document}